\documentclass[twocolumn,preprintnumbers,superscriptaddress,nofootinbib,showpacs,showkeys]{revtex4}
\usepackage{amsmath}
\usepackage{axodraw4j}
\usepackage{amssymb}
\usepackage{graphicx}
\usepackage{hyperref}
\usepackage{xspace}
\usepackage{bm}
\usepackage{slashed}
\usepackage{dsfont}
\usepackage{bbm}

\usepackage{booktabs}
\AtBeginDocument{
\heavyrulewidth=.08em
\lightrulewidth=.05em
\cmidrulewidth=.03em
\belowrulesep=.65ex
\belowbottomsep=0pt
\aboverulesep=.4ex
\abovetopsep=0pt
\cmidrulesep=\doublerulesep
\cmidrulekern=.5em
\defaultaddspace=.5em
}

\newcommand{\be}{\begin{equation}}
\newcommand{\ee}{\end{equation}}
\newcommand{\bea}{\begin{eqnarray}}
\newcommand{\eea}{\end{eqnarray}}

\newcommand{\bpm}{\begin{pmatrix}}
\newcommand{\epm}{\end{pmatrix}}

\usepackage{xcolor}
\definecolor{light-gray}{gray}{0.8}

\begin{document}

\title{Tetraquarks  in the $1/N$ expansion and meson-meson resonances}


\newcommand{\CERNaff}{CERN Theory Department, CH-1211 Geneva 23, Switzerland}
\newcommand{\sapienza}{Dipartimento di Fisica and INFN, `Sapienza' Universit\`a di Roma\\
P.le Aldo Moro 5, I-00185 Roma, Italy}
\newcommand{\alice}{ALICE\xspace}

\author{L. Maiani }
\affiliation{\sapienza}
\author{A.D.~Polosa}
\affiliation{\sapienza}
\affiliation{\CERNaff}
\author{V.~Riquer}
\affiliation{\sapienza}

\begin{abstract}
Diquarks  are found to have the right degrees of freedom to describe the tetraquark poles in  hidden-charm to open-charm meson-meson amplitudes. Compact tetraquarks result as intermediate states in non-planar diagrams of the $1/N$ expansion and the corresponding resonances are narrower than what estimated before.  The proximity of tetraquarks to meson-thresholds has an apparent role in this analysis and, in the language of meson molecules, an halving rule in the counting of states is obtained.

 \end{abstract}


\maketitle

\section{Introduction}
 
The series of $X, Y, Z$ resonances, initiated {\color{black} by Belle} with the $X(3872)$~\cite{x3872} resonance later confirmed by BaBar~\cite{babarx}, CDF~\cite{cdf}, D0~\cite{d0}, LHCb~\cite{lhcbx}, CMS~\cite{cmsx}, has recently been enlarged by the observation of two, hidden charm, pentaquarks $\mathds{P}$ discovered by LHCb~\cite{lhcbpenta}. The $X,Y,Z,\mathds{P}$, hadrons have received different interpretations, under the names of tetraquarks/pentaquarks~\cite{noi1,noi2}, molecules or resonances~\cite{molecules&res} or cusp effects of different kinds~\cite{bibliocusp}. 

To be sure, to explain the exotic hadrons, nobody has challenged the validity of Quantum Chromodynamics or has invoked the presence of new types of fundamental constituents. 
Rather, the existence of different pictures seems to reflect our  ignorance about the exact solutions of non-perturbative QCD. 
Different interpretations call into play different approximations or different regimes of the basic QCD force, to arrive to seemingly contradictory pictures. 

The tetraquark and pentaquark description utilizes as a guiding framework the non-relativistic quark constituent model, which has given an accurate picture of  $q\bar q$ and $qqq$ mesons and baryons, including charmed and beauty hadrons. The starting point is the attraction within a color antisymmetric quark pair, which arises in perturbative QCD due to one-gluon exchange and in non-perturbative QCD due to instantons~\cite{instant}. This makes diquarks and antidiquarks suitable basic units to build $X, Y, Z$ and pentaquark hadrons, with mass splittings due to spin-spin interactions and orbital momentum excitation~\cite{noi1,noi2}.

Some molecular models assume $X, Y, Z$ peaks to be produced by color singlet exchange forces between $q \bar q$ color singlet mesons, or to be produced by kinematic singularities due to triangle diagrams and the like~\cite{molecules&res}.

Thus far, no `smoking gun' signature has been found to distinguish the different interpretations, with the exception of the naive, loosely bound, molecular model, largely unfavored by the production cross section of $X(3872)$ at large $p_{T}$ in high energy hadron colliders~\cite{prod}; for a compelling comparison with some recent Alice data see~\cite{prod2}. 
Perhaps, these models are, after all, different but complementary descriptions of the same QCD underlying reality. 

In this paper we use the $1/N$ expansion of $N$-colors QCD~\cite{ninfty} to investigate the relations between the exotic meson description in terms of diquark-antidiquark bound states and  of resonances in meson-meson scattering. 

Following~\cite{weinberg,knecht},  we consider as basic units of tetraquarks the natural extension to $N$ colors of antisymmetric diquark operators. 
 It was assumed in {\color{black} Refs.~\cite{weinberg,knecht} that tetraquark correlation functions could develop poles at the level of planar diagrams,  of order $1/N$ with respect to the leading, disconnected, amplitudes. We argue this to be an unlikely occurrence and explore the possibility that genuine tetraquark poles arise to higher orders in the $1/N$ expansion.} 
In the present work, considering the correlation of charged tetraquarks, we find a picture consistent with the factorization of the amplitudes at the poles, at the level of  {\it correlation functions with  one handle}. 

We consider further the correlators of neutral tetraquarks, that may mix with genuine charmonia. We find consistency for the mixing constant derived from tetraquark correlators and from two-point meson-meson  correlators, where tetraquarks appear via internal quark loops -- see Fig.~\ref{mixing} -- resolving an inconsistency which appeared in Ref.~\cite{knecht}.

Next we consider meson-meson scattering amplitudes which are generated, at order $1/N$, by one quark loop with four external color singlet sources. Choosing appropriate external quark flavours and assuming that the annihilation of heavy quarks is suppressed, we construct amplitudes which can receive contributions from four quark intermediate states only. In a simplified model, we find that meson-meson scattering eigenchannels coincide with the color singlet states obtained by  Fierz rearrangement of symmetyric or antisymmetric diquark-antidiquark operators. If the antisymmetric tetraquark correlator develops a pole, this would appear as an exotic resonance in meson-meson scattering.  In the language of meson molecules, an halving rule in the counting of states is obtained, and the proximity of tetraquarks to meson-thresholds has an apparent role in this analysis.

With poles of order $1/N^3$ with respect to the meson meson correlation found in~\cite{coleman}, one may argue that tetraquarks exist but are largely decoupled from the meson sector. The question is answered by Weinberg's analysis~\cite{weinberg}. Not only it is important that tetraquark poles might appear in the diagrams of the $1/N$ expansion, but also that the related resonances are narrow enough to be distinguishable from background. Obviously at $N=\infty$ the coupling is zero and tetraquarks are never produced. However, we know that this strict limit has little to do with phenomenology: at $N=\infty$ mesons are free particles. In conclusion, if tetraquark poles are formed, and this is admittedly a big if, the most relevant question is if the total widths are $\Gamma\sim N^\alpha$ with $\alpha <0$, which is answered positively in Weinberg's analysis. 

Even in the new framework, we confirm that tetraquark widths decrease as $N\to\infty$.

Tetraquarks in the $1/N$ expansion have been considered in a number of papers with different approaches.

 In~\cite{Cohen:2014via} quarks are described by the antisymmetric representation for $N\geq 3$, an extreme version of the Corrigan-Ramond scheme~\cite{Corrigan:1979xf}.~
 What found in~\cite{Cohen:2014via} is that in the large-$N$ limit  one can produce tetraquarks in a completely natural way, because new color-entangled operators exist.
 
 In~\cite{Cohen:2014tga} it is shown that the Coleman-Witten lore that no tetraquarks occur at large $N$ is not related to the fact that they did not consider all possible ways of cutting the diagrams.  Therefore, tetraquarks can be made the way suggested in~\cite{weinberg}, but with the concerns discussed in~\cite{Lebed:2013aka}. 

The fact that subleading topologies may be important, as discussed later in our paper, seems to emerge also to explain the large $N$ behavior of the lightest scalars~\cite{Cohen:2014vta}.

{\color{black}
\section{Diquarks and tetraquarks in SU(N)}
Consider two quarks interacting through the exchange of one virtual gluon in $N=3$ QCD as in Fig.~\ref{dq} (the case of a quark-antiquark pair, in connection with singlet confinement, was originally considered by Han and Nambu~\cite{hannambu}, the following considerations can be repeated for antiquarks).
\begin{figure}[htb!]
 \begin{center}
   \includegraphics[width=0.65\columnwidth]{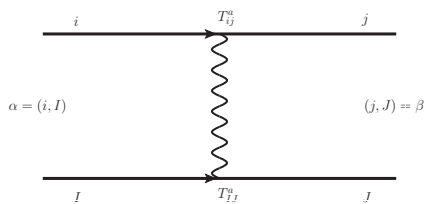}
 \end{center}
\caption{\footnotesize One-gluon exchange interaction.}   
\label{dq}
\end{figure}

The $T^a_{ij}T^a_{IJ}$ tensor product can be mapped into a $9\times 9$ matrix whose entries $A_{\alpha\beta}$ correspond to the 81 possible combinations of initial and final colors as in Fig.~\ref{dq}. The $v$ eigenvectors of $A$  identify  3  antisymmetric color configurations and 6 symmetric ones.  
For each $v$ the  $v^T A v$ product is a  superposition of the color diagrams in  Fig.~\ref{dq} 
defining amplitudes which are (anti)-symmetric under the simultaneous exchange of the colors $i\to I, j\to J$.}

Each of these 9 color configurations is  weighted by a coefficient $h$, the  eigenvalue related to $v$. The $h$ are found to be negative in the antisymmetric cases and positive in the symmetric ones: $h=-2/3$ and $h=1/3$ respectively for SU(3). 
The value of $h$ corresponds to the product of charges in a abelian theory -- thus one gets repulsion in the symmetric eigenchannels~\footnote{e.g. for $i=I,j=J$.} and attraction in the antisymmetric ones. 

The eigenvalues $h$ are more conveniently computed through the quadratic Casimirs of the irreducible representations $\bm S_i$ obtained from the Kronecker decomposition of the product $\bm R_1\otimes \bm R_2=\bm S_1\oplus \bm S_2\oplus ...$. In the case of quark-quark interaction in SU(3), $\bm R_1= \bm R_2=\bm 3$ and $\bm S_1=\bar{\bm 3}$, $\bm S_2=\bm 6$. The formula for the eigenvalues $h_i$ in the various eigenchannels is in general
\be
h_{i}=\frac{1}{2}(C_{\bm S_i}-C_{\bm R_1}-C_{\bm R_2})
\ee
where $C_{\bm S_i}, C_{\bm R_1}, C_{\bm R_2}$ are the quadratic Casimirs in the 
$\bm S_i, \bm R_1, \bm R_2$ representations respectively.

In the generic case of SU(N) we have that 
\be
\bm N\otimes \bm N=\bm{\frac{N(N-1)}{2}}\oplus \bm{\frac{N(N+1)}{2}}
\ee
where $\bm{N(N-1)/2}$ is antisymmetric and $\bm{N(N+1)/2}$ is symmetric. 
 
 The Casimirs associated to these representations are given in the following table

 \begin{table}[htb!]
\centering
    \begin{tabular}{||c|c|c||} \hline
{\footnotesize Representation $\bm R$}  & {\footnotesize $C_{\bm R}$} &  {\footnotesize $h$ }\\ \hline
 {\footnotesize $\bm N$} & {\footnotesize $(N^2-1)/(2N)$} & $-$ \\ 
\hline
{\footnotesize $\bm{N(N+1)/2}$} & {\footnotesize $(N-1)(N+2)/N$} &{\footnotesize  $(N-1)/2N>0$}\\
\hline
{\footnotesize $\bm{N(N-1)/2}$} & {\footnotesize $(N+1)(N-2)/N$} & {\footnotesize $-(N+1)/2N<0$}\\
\hline\hline
\end{tabular}
 \caption{\footnotesize Quadratic Casimir operators for the fundamental, the two index symmetric and antisymmetric representations, in color SU(N), $N\geq 2 $. In the third column, the coefficient of the potential energy for color symmetric and antisymmetric diquarks in the one-gluon exchange approximation. Attraction in the antisymmetric channel persists at large $N$.}
\label{casimir}
\end{table}

In the singlet channel of $\bm N\otimes \overline{\bm N}$, the attraction is weighted by $h=-(N^2-1)/2N$. Therefore, the singlet channel is $(N-1)$ more attractive than the antisymmetric 
$\bm{N(N-1)/2}$ channel reported in Tab.~\ref{casimir}, in SU(3). In the one-gluon exchange approximation, the singlet channel in $q\bar q$ is (just) twice more attractive, for $N=3$, than the color antitriplet channel in $qq$.

For any value,  $N$, diquark operators for two quarks with given flavors $q,q^\prime$ can be  written in symmetric $(S)$ or antisymmetric $(A)$  color configurations
\be
 {d}^{S,A}_\Gamma=q^\alpha \Gamma q^{\prime \beta}  \pm q^\beta \Gamma q^{\prime \alpha}
\label{diqop}
\ee
with $\Gamma$ matrices to characterize the diquark spin and $\alpha,\beta$ color indices.

Color forces may bind a diquark-antidiquark pair in a tetraquark, the analog of usual mesons with the substitutions 
\bea 
&&q \to {\bar d}^A  \label{sosq}\\
&&\bar q \to d^A \label{sosqbar}
\eea

In $N=3$ QCD, there is a special relation between baryons and tetraquarks. 
If we start from an antibaryon, the substitution in Eq.~(\ref{sosqbar}) produces the tetraquark $d^A \bar q \bar q= d^A{\bar d}^A$. Applying~(\ref{sosqbar}) once again, one obtains a {\it pentaquark}, $d^A d^A\bar q$, and finally, with a third substitution, a state with baryon number $B=2$, a {\it dibaryon} with the configuration $d^A d^A d^A$. 

This chain of reasoning motivates the alternative generalization of  $N=3$ tetraquarks to arbitrary $N$ proposed by G. C. Rossi and G. Veneziano~\cite{randv,randv2}. The diquark in Eq.~(\ref{diqop}) is generalized to the fully antisymmetric product of $N-1$ quark fields
\be
{\cal M}_\alpha=\epsilon_{\alpha \beta_1 \beta_2 \cdots \beta_{N-1}} q^{\beta_1} q^{\beta_2} \cdots q^{\beta_{N-1}}
\label{multiq}
\ee
and hadrons can be formed as color singlet combinations  ${\cal M} {\bar{\cal M}}$. For excited multiquark hadrons, the color string connecting ${\cal M}$ to ${\bar {\cal M}}$ can break with production of a baryon-antibaryon pair
\be
({\cal M} {\bar {\cal M}}) \to B\, \bar B
\ee
${\cal M} {\bar {\cal M}}$ hadrons below the baryon antibaryon threshold would be narrow, whence the name {\it baryonium} given to these mesons.

The treatment of baryons in the large $N$ limit was initiated by Witten~\cite{witten} and several works followed~\cite{Dashen:1993jt,Dashen:1994qi,Luty:1993fu,Carone:1993dz} including discussions on the excited baryons as in~\cite{Carlson:1998gw}.

{\color{black} We restrict in the following to the generalization embodied in Eq.~(\ref{diqop}), using the diquark fields $d^A$ to construct interpolating operators which create or annihilate tetraquarks for any $N$.}

\section{Fierz rearrangement}

We restrict  to hidden charm tetraquarks and focus, at first,  on charged, isospin $I=1$ tetraquarks. Neutral tetraquarks will in general mix with charmonium resonances and have to be considered separately.

Simple but sufficiently representative examples  of hidden charm tetraquarks are given by the 
$S$-wave states with $J^P=1^+$ and isospin $I=1$, which can be classified according to 
G-parity
\be
G={\cal C} e^{i\pi I_2}
\ee
that is charge-conjugation, ${\cal C}$, accompanied by a $180^0$ rotation in isospin space which brings $I_3 \to -I_3$.

Following~\cite{noi2} $X^+$ denotes the predicted, but not (yet) observed, charged counterpart of $X(3872)$~\footnote{$X^+$ could  be very broad, as suggested in~\cite{tera}, or be altogether suppressed by the mechanism suggested in~\cite{fere}.}, and we identify $Z^+=Z^+(3900)$~\cite{zzdisc}, $Z^\prime=Z^+(4020)$~\cite{4020}. We report first the explicit formulae for antisymmetric diquarks, with $q$ and $\bar  q$ the two-dimensional spinors representing the annihilation operators for quarks and for the charge conjugate antiquarks $\bar u, \bar d, \bar c$
\bea
&& \underline{G=-1}:\nonumber\\
&&X^+=(c^\alpha \sigma^2 u^\beta)\left[({\bar c}_\alpha \sigma^2{\bm \sigma} {\bar d}_\beta)-(\bar c_\beta \sigma^2{\bm \sigma} {\bar d}_\alpha)\right]+\nonumber \\ 
&& \quad\quad +\,  (\sigma^2 \leftrightarrow \sigma^2{\bm \sigma})\label{xx}\\
&& \underline{G=+1}:\nonumber \\
&& Z^+=(c^\alpha \sigma^2 u^\beta)\left[({\bar c}_\alpha \sigma^2 {\bm \sigma}{\bar d}_\beta)-(\bar c_\beta \sigma^2 {\bm \sigma} {\bar d}_\alpha)\right] - \nonumber \\ 
&&\quad\quad -\,  (\sigma^2 \leftrightarrow \sigma^2{\bm \sigma})\label{zz}\\
&&Z^{\prime +}=(c^\alpha \sigma^2{\bm \sigma} u^\beta)\wedge \left[({\bar c}_\alpha \sigma^2{\bm \sigma} {\bar d}_\beta)-({\bar c}_\beta \sigma^2{\bm \sigma} {\bar d}_\alpha)\right]\label{zzp}
\eea
where $\sigma$ denote Pauli matrices. In the first and following lines we have used the antisimmetrization of the diquark colors, Eq.~(\ref{diqop}), $\sigma^2$ and $\sigma^2{\bm \sigma}$, project quark-quark or antiquark-antiquark bilinears with spin $0$ and spin $1$, respectively.

\begin{table}[htb!]
\centering
    \begin{tabular}{||c|c|c|c||} \hline
-- &{\footnotesize$\begin{array}{c}(\sigma^2)_{ik}\\ (\sigma^2{\bm \sigma})_{jl} \end{array}$}& {\footnotesize $\begin{array}{c}(\sigma^2{\bm \sigma})_{ik}\\(\sigma^2)_{jl}\end{array}$} &  {\footnotesize $\begin{array}{c} i (\sigma^2{\bm \sigma})_{ik}\wedge \\ \wedge (\sigma^2{\bm \sigma})_{jl}\end{array}$ }\\ \hline
 {\footnotesize $(\sigma^2)_{ij}(\sigma^2{\bm \sigma)}_{kl} $} &{\footnotesize $+1/2$} & {\footnotesize $-1/2$} & {\footnotesize $+1/2$} \\ 
\hline
{\footnotesize $(\sigma^2{\bm \sigma})_{ij}(\sigma^2)_{kl} $} &{\footnotesize $-1/2$} & {\footnotesize $+1/2$} & {\footnotesize $+1/2$} \\
\hline
{\footnotesize $i(\sigma^2{\bm \sigma})_{ij}\wedge (\sigma^2{\bm \sigma})_{kl}$} &{\footnotesize $+1$} & {\footnotesize $+1$} & {\footnotesize $0$}\\
\hline
\end{tabular}
 \caption{\footnotesize Coefficients for the Fierz rearrangment of $J^P=1^+$ quadrilinears.}
\label{fierz}
\end{table}

Products of two diquark operators can be expressed in terms of color singlet bilinears, with coefficients determined by the Fierz-rearrangement coefficients reported in Table~\ref{fierz}.

For $X^+$, Eq.~(\ref{xx}), using Table~\ref{fierz} and  $\sigma^2\bm \sigma=(\sigma^2\bm \sigma)^T$, $(\sigma^2)^T=-\sigma^2$, we find
\bea
&&X^+=-i(c\sigma^2 {\bm \sigma}{\bar c})\wedge (u\sigma^2{\bm \sigma}{\bar d})-\nonumber \\
&&- \left[c\sigma^2{\bar d})(u\sigma^2{\bm \sigma}{\bar c}) - (c\sigma^2{\bm \sigma}{\bar d})(u\sigma^2{\bar c})\right]
\label{colsing}
\eea

With symmetric diquarks, we would get a plus sign inside the brackets of Eq.~(\ref{xx}) or (\ref{colsing}).
We can also formally identify color singlet bilinears with the $S$-wave mesons, and, up to an overall normalization, write
\be
X^+ \sim i\frac{\psi \wedge \rho^+}{\sqrt{2}} \pm \frac{{\bar D}^0 D^{\star +}-D^+ {\bar D}^{\star 0}}{\sqrt{2}} 
\label{xxmesons}
\ee
normalized Pauli bilinears and the plus/minus sign is for symmetric/antisymmetric diquarks.

Similarly,  for the other operators we have
\bea
Z^+&=&-\left[(c\sigma^2 {\bar c})(u\sigma^2{\bm \sigma}{\bar d})-(c\sigma^2{\bm \sigma} {\bar c})(u\sigma^2{\bar d})\right] \pm \nonumber\\
&\pm& i (c\sigma^2 {\bm \sigma}{\bar d})\wedge (u\sigma^2{\bm \sigma}{\bar c})=\nonumber \\
&\sim& \frac{\eta_c \,\rho^+ - \psi\, \pi^+}{\sqrt{2}} \pm i \frac{{\bar D}^{0 \star} \wedge D^{\star +}}{\sqrt{2}}
\label{zzmesons}
\eea
and 
\bea
Z^{\prime +}&=&-\left[(c\sigma^2{\bar c})(u\sigma^2{\bar d})+(c\sigma^2{\bm \sigma} {\bar c})(u\sigma^2{\bar d})\right]\mp \nonumber  \\
& \mp& \left[ (c\sigma^2 {\bar d})(u\sigma^2{\bm \sigma} {\bar c})+(c\sigma^2 {\bm \sigma}{\bar d})(u\sigma^2 {\bar c})\right]=\nonumber \\
&\sim& \frac{\eta_c \,\rho^+ +\psi \,\pi^+}{\sqrt{2}}\pm \frac{{\bar D}^0 D^{\star +} + D^+ {\bar D}^{ \star 0}}{\sqrt{2}}
\label{zzpmesons}
\eea 
We will come back to these formulae in Sections~IV and~VI.

We remark that the labels $Z$ and $Z^\prime$ chosen for the two states $Z(3900)$ and $Z(4020)$, are the same as those used in~\cite{noi2} where it is shown that the heavier state has both diquarks in spin 1 whereas the lighter has one diquark in spin 0 and one in spin 1, see Eqs.~(\ref{zz},\ref{zzp}).   The Fierz rearrangement of the quarks, if hadronization effects are not taken into account, would predict the kinematically forbidden $Z\to \bar D^* D^*$ decay, see Eq.~(\ref{zzmesons}) instead of ${\bar D}D^*$, as indicated by the observed $Z$ decay, and the opposite is exposed in Eq.~(\ref{zzpmesons}). However formulae in Eqs.~(\ref{zzmesons}) and~(\ref{zzpmesons}) do not take into account the effects of hadronization, which might affect the spin of the light quarks leaving the heavy quark pair spin unchanged~\cite{alib}. Hadronization might produce a $\bar DD^*$ state in~Eq.~(\ref{zzmesons}), or add an~$h_c\pi^+$  to  the $\eta_c\rho^+$component. ~
As a consequence, we would expect that $Z^\prime$ has a ${\bar D}D^*$ decay in addition to the observed ${\bar D}^*D^*$ mode, an interesting point to check experimentally.

\begin{figure}[htb!]
 \begin{center}
   \includegraphics[width=0.55\columnwidth]{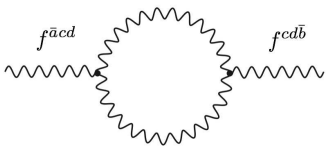}
 \end{center}
\caption{\footnotesize Gluon self-energy diagram with fixed colors $\bar a=\bar b$.}   
\label{gb}
\end{figure}

{\color{black}
\section{Large N expansion: a reminder}

The behavior of QCD for $N\to \infty$ has been characterized by G.'t Hooft~\cite{ninfty}.} Consider the gluon self-energy diagram in Fig.~\ref{gb}, with gluon colors fixed to $\bar a,\bar b$.
This diagram involves the product
\be
\sum_{c,d}f^{\bar a cd} f^{\bar b cd}=\mathrm{Tr}(T^{\bar a}T^{\bar b})=N\,\delta^{\bar a\bar b}
\ee
in the adjoint of SU(N). The gluon loop therefore contains a multiplicity factor of $N$ in SU(N). 

{\color{black} To make the large $N$ limit of this diagram smooth, one requires that the couplings at vertices, $g_{_{\rm QCD}}$, scale with $N$ as $g_{_{\rm QCD}}=g_c/\sqrt{N}$ so that 
\be
\frac{g^2_c}{N}\times N=g^2_c~\text{independent of $N$}
\ee 
Sometimes one refers to the 't Hooft coupling \
\be
\lambda=g_{_{\rm QCD}}^2 N
\ee
 ($g_c=\sqrt{\lambda}$ in this notation). The large-$N$ limit is obtained keeping $\lambda$ fixed. 

The gluon field is characterized by the color indices
\be
(A_\mu)^i_j=(T^a)^i_j\, A^a_\mu
\ee
The number of independent components of this matrix in SU(N) are $N^2-1$. In the large $N$ limit however we can treat it as a $N\times N$ matrix and represent the gluon line by a double color line -- carrying a pair of color indices $i,j$.
With this notation the diagram in Fig.~\ref{gb} can be represented as in Fig.~\ref{gbl}.
\begin{figure}[htb!]
 \begin{center}
   \includegraphics[width=0.5\columnwidth]{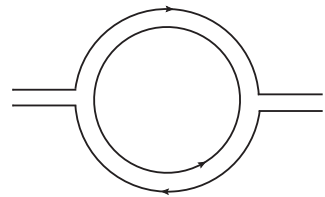}
 \end{center}
\caption{\footnotesize Gluon self-energy diagram in the large $N$. With this notation the multiplicity factor  $N$ traced above in the $f$ structure constants, has a clear origin in the color loop at the center.}   
\label{gbl}
\end{figure}

The origin of the multiplicity factor discussed above becomes apparent in the double-line notation. The quark-gluon coupling will therefore also scale as $g_c/\sqrt{N}$ and the 4-linear gluon coupling as $g^2_c/N$.

't Hooft shows that in the  $N\to \infty$  limit only planar diagrams with quarks along the external edge survive. 

The rule is easily visualized by computing the correlation function of a color singlet quark bilinear with itself. With no gluon lines, the result is obviously proportional to $N$, the number of colors that run in the loop. A gluon line traversing the loop, see Fig.~\ref{loopuno}, can be represented by two color lines running in opposite directions and joining the quark and antiquark lines that flow in the vertex. 
Thus we get two loops, {\it i.e.} a factor of $N^2$, times the factor $g_c^2/N$, therefore a contribution of order $\lambda \times N$. 

The sum of all planar diagrams of this kind will again be of order $N$, times a non-perturbative function of $\lambda$, which may well develop poles for certain values of the external momentum, $q^2$.

  \begin{figure}
 \begin{center}
   \includegraphics[width=1.0\columnwidth]{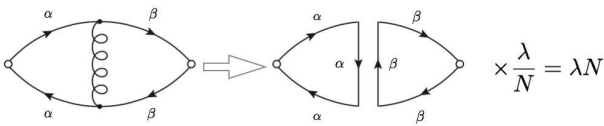}
 \end{center}
\caption{\footnotesize One gluon exchange correction to the correlation function of a color singlet quark bilinear, represented by the open circle. Representing  the gluon line by two, oppositely running lines joining the quark lines on the edge, one sees that the diagram reduces, for color number counting, to a two loop diagram. Thus one recovers a result of order $N$, like the lowest order diagram, multiplied by the color reduced coupling $\lambda		$. }   
\label{loopuno}
\end{figure}

The sum of all planar diagrams like the one on the lhs of Fig.~\ref{loopuno} is represented by
\be
\langle 0|J(p)J^*(p)|0\rangle\sim N
\ee
where the operator $J^*$ acts on the vacuum to create a meson state, and
 \bea
\langle 0|J(p)J^*(p)|0\rangle&=&\sum_n \frac{\langle 0|J(p)|n\rangle\langle n|J^*(p)|0\rangle}{p^2-m_n^2}=\notag\\
&=&\sum_n \frac{f_n^2}{p^2-m_n^2}
\label{spettrale}
\eea
with the decay constant $f_n=\langle 0|J(p)|n\rangle$.

The behavior at large $p^2$ momenta of  $\langle 0|J(p)J^*(p)|0\rangle$ is expected to be logarithmic and the sum over meson states can behave as $~\sim \ln p^2$, at large $p^2$, only if it has an infinite number of terms, as can be seen by  $\sum_n\to \int dm_n^2$. Thus we have an infinite number of poles, corresponding to a tower of (stable) meson states in the correlation function  $\langle 0|J(p)J^*(p)|0\rangle$. These have a given flavor content, {\it e.g.} quarkonium mesons with varying quantum numbers, like quark spin and orbital angular momenta, radial excitations, {\it etc}.
Meson masses are independent of $N$ and the entire $N$ dependency of the lhs of~(\ref{spettrale}) is encoded in $f_n$. In the case at hand this means\footnote{ actually $f_n\sim \sqrt{N}(1+a/N+b/N^2+...)$. } that each $f_n\sim\sqrt{N}$.

{\color{black} It may be convenient to extract a factor of $\sqrt{N}$ from each $f_n$ to obtain a propagator which is $N$- independent in the large $N$ limit.}
Equation~(\ref{spettrale}) can then be written graphically in the meson theory as
\vspace{-30pt} \hfill \\ 
\SetScale{0.3} \SetWidth{2}
\begin{center}
$\langle 0|J(p)J^*(p)|0\rangle=~\sum_n~~~~$
\begin{picture}(70,30)(0,13) 
 \Text(-8,17)[]{$\sqrt{N}$}  \Text(38,17)[]{$\sqrt{N}$} \BCirc(20,50){5}{}{}\Line(26,50)(80,50) \BCirc(80,50){5}{}{}  \Text(15,20)[]{\small $n$}
\end{picture}
\end{center}
where open dots indicate the decay constants $f_n$ normalized so as to have a finite limit for $N\to\infty$.


\section{Tetraquark correlators in the large N expansion}

One may consider correlation functions of tetraquark operators, like those given in Eqs.~(\ref{xx}) to (\ref{zzp}) and see if they develop poles, as is the case for  $q\bar q$ operators.  

The reputation of tetraquarks was somehow obscured by a theorem of S. Coleman~\cite{coleman} stating that: {\it tetraquarks correlators for $N \to \infty$ reduce to disconnected meson-meson propagators}. 

The theorem follows from the simple fact that a four quark operator can be reduced to products of color singlet bilinears, see Eqs.~(\ref{colsing}) to (\ref{zzpmesons}).  Connecting each bilinear with itself, one gets two disconnected one-loop diagrams, {\it i.e.} a result of order $N^2$, while connected tetraquark diagrams are one-loop,  see Fig.~\ref{loop5}, thus  of order $N$.

The argument was reexamined by S.~Weinberg~\cite{weinberg} who argued that {\it if connected tetraquark correlators develop a pole, it will be irrelevant that its residue is of order 1/N with respect to the disconnected parts}. After all, meson-meson scattering amplitudes are of order $1/N$, in the $N\to \infty$ limit, and we do not consider mesons to be really free particles.

The real issue, according to Weinberg, is the width of the tetraquark pole: if it increases for large $N$, the state will be undetectable for $N\to \infty$.
Weinberg finds that decay rates go like $1/N$, making tetraquarks a respectable possibility. 
The discussion has been enlarged by M.~Knecht and S.~Peris~\cite{knecht} and by R. Lebed~\cite{Lebed:2013aka}, see also~\cite{Esposito}.

  \begin{figure}
 \begin{center}
   \includegraphics[width=0.95\columnwidth]{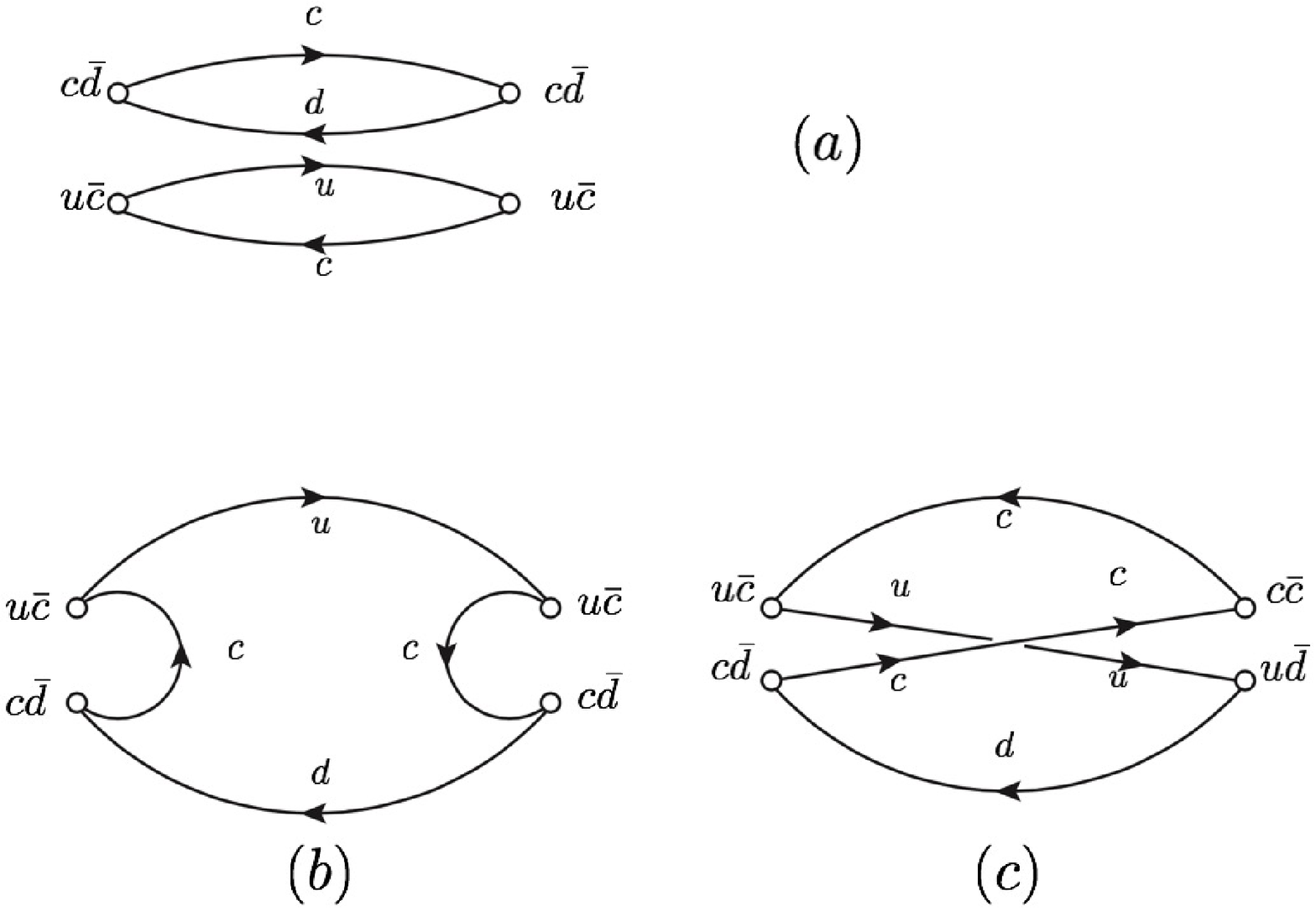}
 \end{center}
\caption{\footnotesize Connected and disconnected diagrams for the tetraquark correlation functions: $(c\bar d)(u\bar c)\to (c\bar d)(u\bar c)$ and $(c\bar d)(u\bar c)\to (c\bar c)(u\bar d)$.  }   
\label{loop5}
\end{figure}

For the charged tetraquark operators considered in Eqs.~(\ref{xx}) to (\ref{zzp}), there are only two connected tetraquark correlators shown in Fig.~\ref{loop5}, see the discussion in~\cite{knecht}. Note, however, that diagram (b) corresponds the annihilation of a $c \bar c$ pair to produce a $u\bar d$ intermediate state, which is unfavoured for $m_c \gg \Lambda_{_{\rm QCD}}$, so we remain with diagram (c), which switches a pair of color singlet, open charm, operators into a pair made of one hidden charm and one charmless operator and viceversa.

The situation is different for hidden charm, electrically neutral tetraquarks, of the form, {\it e.g.}, $[cu][\bar c \bar u]$, in that there is an additional diagram analogous to Fig.~\ref{loop5}~(b), with $u\bar u$ annihilating to produce an intermediate charmonium $c \bar c$ state. This diagram, which is not suppressed, produces a $[cu][\bar c \bar u] \leftrightarrow (c\bar c)$ mixing, as discussed in~\cite{knecht} and in Sect.~\ref{neutral}. 

\section{Need of non-planar diagrams}

In previous discussions, it was considered, implicitly~\cite{weinberg} or explicitly~\cite{knecht}, that the diagrams in Fig.~\ref{loop5} (b) and (c) may develop a tetraquark pole of order $N$, namely at the level of planar diagrams. At a closer inspection, this seems to be rather unlikely for the following reason.

Consider the diagram in Fig.~\ref{loop5}~(c). All its cuts contain a two quark-two antiquark state. However such states correspond to two non interacting meson states more than a tetraquark closely bound by color forces. This is even more evident in Fig.~\ref{loop5} (b) which, cut vertically to produce a tetraquark state, gives, in the planar approximation, precisely two non-interacting meson states, of the kind produced by cutting Fig.~\ref{loop5} (a).

These considerations agree with Witten's conclusion~\cite{witten} that in the planar diagram approximation and with reference to Fig. ~\ref{loop5} (c),  the meson-meson scattering, $M_1+M_2 \to M_3 + M_4$ amplitudes has only meson poles in the $u$  and $t$ ($M_1 M_4$ and $M_1 M_3$) channels and no singularity in the $s$ channel ($M_1 M_2$), which is precisely where the tetraquark pole should appear. 

On the other hand, adding non-planar gluons in Fig. ~\ref{loop5} (c), we may connect the tips associated to the insertions of $u\bar c$ and $c\bar d$ and produce the interaction needed to make the color singlet bilinear to merge in a tetraquark. An example is given in Fig.~\ref{nonplanar}.
Nonplanar gluons evade the separation of the diagrams into non interacting meson pairs by unitarity cuts.

Following 't Hooft~\cite{ninfty}, general diagrams can be classified according to the number of quark loops,  $L$, and the number of handles, $H$, the general order in $N$ of the diagram being
\footnote{The gauge theory of large-$N$ QCD is $SU(N)$, which differs from the $U(N)$ assumed in 't-Hooft analysis by a spurious $U(1)$ gluon. One may wonder if, removing  the spurious U(1) gluon from the U(N) theory in order to have SU(N), the expansion in loop and handles in~(\ref{handle}) remains valid.
We cannot answer this question except that we do not see how a color neutral gluon can eventually modify the $N$ counting.
} $N^\alpha$ with
\be
\alpha= 2- L-2H
\label{handle}
\ee
The diagrams in Fig.~\ref{loop5} (b), (c) have $L=1$ thus are of order $N$ for $H=0$. The diagram reported in Fig.~\ref{nonplanar} has $L=1, H=1$ and therefore is of order $N^{-1}$, as can be directly verified with the rules given previously. To proceed further, we have to distinguish the two cases of charged, isospin one, and neutral, isospin zero, tetraquarks
\paragraph*{\bf Charged tetraquarks.}

 For simplicity we assume that one handle as in Fig.~\ref{nonplanar}, is sufficient to develop the charged, isospin one, tetraquark pole. The previous symbolic equation is therefore replaced by
\vspace{-30pt} \hfill \\ 
\SetScale{0.3} \SetWidth{2}
\begin{center}
$\langle 0|{\cal Q}(p){\cal Q}^*(p)|0\rangle=\sum_n~~~~~$
\begin{picture}(70,30)(0,13) 
\Text(-8,17)[]{$\frac{1}{\sqrt{N}}$}  \Text(38,17)[]{$\frac{1}{\sqrt{N}}$} \CCirc(21,50){5}{}{}\Line[double,sep=5](26,50)(80,50) \CCirc(80,50){5}{}{} \Text(15,20)[]{\small $n$}
\end{picture}
\end{center}
where ${\cal Q}$ are $I=1$ tetraquark operators and full dots indicate the corresponding constants $f_n$, normalised to have a finite limit for large $N$.

  \begin{figure}
 \begin{center}
   \includegraphics[width=0.95\columnwidth]{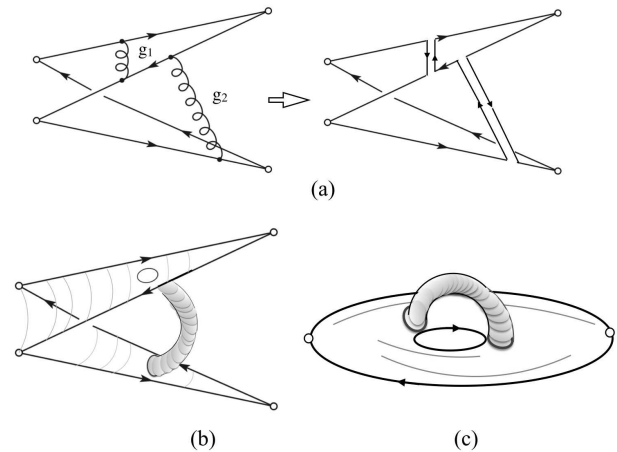}
 \end{center}
\caption{\footnotesize  (a) The simplest non-planar modification of the diagram in Fig.~\ref{loop5} (c). If there were the planar gluon, $g_1$, only, the representation of a gluon line with two color lines would generate two color loops and the corresponding amplitude, multiplied by the coupling factor $\lambda/N$, would be again of order $N$. The non planar gluon, $g_2$, makes a bridge between the two color loops, bringing back to one loop only. The amplitude, multiplied by the coupling factor $\lambda^2/N^2$ is therefore of order $1/N$, as expected for a diagram with one quark loop and one handle Eq.~(\ref{handle}). (b), (c) Non perturbative realizations of the one-handle diagram in (a) and of the two quark loop diagrams to be considered later, Fig.~\ref{mixing} (c), (d). }   
\label{nonplanar}
\end{figure}

 Consider now the decay amplitude into two mesons. The correlation functions in Fig.~\ref{loop5} (c) or Fig.~\ref{nonplanar} involve the tetraquark operator and two quark bilinear insertions, represented by the dots in the final state.  

The decay of a tetraquark in the meson theory can be represented by the diagram of Fig.~\ref{decaytqmm}. Since the correlation function is itself of order $1/N$, we find the decay constant
\be
g^{(1)}=\frac{1}{N\sqrt{N}}
\label{decamp1}
\ee
Our result is reduced by a factor $1/N$ with respect to the estimate in Reffs.~\cite{weinberg,knecht}, in correspondence to the introduction of non-planar diagrams. The conclusion that the width of tetraquarks vanishes for $N\to \infty$ still applies.

 \begin{figure}[h!]
 \begin{center} \includegraphics[width=0.55\columnwidth]{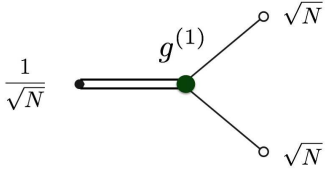}
 \end{center}
\caption{\footnotesize Tetraquark decay in two standard mesons in the meson theory. Its order in $N$ has to match the order of the non-planar diagram in Fig.~\ref{nonplanar}.}   
\label{decaytqmm}
\end{figure}


An alternative but equivalent description comes from assigning a factor $\sqrt{N}$ to each tetraquark insertion and a factor $1/\sqrt{N}$ to each standard meson insertion on the color loop diagrams -- this is done in~Fig.~\ref{decay} (a).  There we have again 
\be
\sqrt{N} \times \left(\frac{1}{\sqrt{N}}\right)^2\times \frac{1}{N}=\frac{1}{N\sqrt{N}}=g^{(1)}
\ee

  \begin{figure}
 \begin{center}
   \includegraphics[width=0.85\columnwidth]{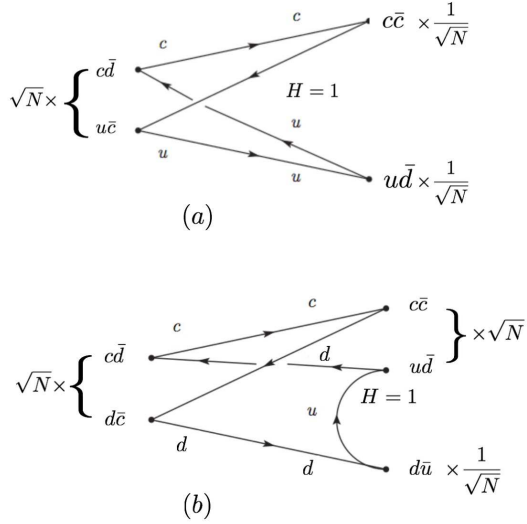}
 \end{center}
\caption{\footnotesize Normalized decay amplitudes:  (a) $Z^+ \to J/\Psi + \pi^+$, (b) $Y (I=1) \to Z^+ + \pi^-$. The amplitude of of quark loops with one handle is of order $1/N$.}   
\label{decay}
\end{figure}

One can also describe the process of tetraquark de-excitation, such as the decay of a $Y$ state into $S$-wave diquarks with the emission of a pion,  $Y(4260) \to Z^+ + \pi^-$, Fig.~\ref{decay} (b). In this case, there are two normalization factors for the tetraquarks  and one for the meson , giving
\be
\left(\sqrt{N}\right)^2\times \frac{1}{\sqrt{N}}\times \frac{1}{N}=\frac{1}{\sqrt{N}}=g^{(1)} \times N
\ee

Replacing the meson insertion with the electromagnetic current, one obtains the amplitude for radiative decays, such as $Y(4260) \to X(3872) +\gamma$, reported in~\cite{Ablikim:2013dyn}. There is no $1/\sqrt{N}$ normalization factor for the current and radiative decay rates are of order $N\alpha$, with respect to de-excitation in a meson.

\paragraph*{\bf Neutral, $I=0$, tetraquarks.}\label{neutral}

Mixing of neutral, $I=0$ tetraquarks with charmonia has been considered in~\cite{knecht} for the case of planar diagrams. Here we consider the realistic case where the tetraquark pole arises at the level of non-planar diagrams, assuming the smallest number of handles allowed by factorization of the correlation functions. 

The simplest case is given in  Fig.~\ref{mixing} (a), which relates the amplitude of the one-handle diagram (a) to the product
\be
f_{4q}\, f
\label{prodf}
\ee
Observe that as a consequence of the non-planar topology, a gluon, rather than the quark, performs part of the periphery and the diagram cannot be deformed in a two-point meson correlation function without changing its order in the $N$ power  counting.
  \begin{figure}[htb!]
 \begin{center}
   \includegraphics[width=1.0\columnwidth]{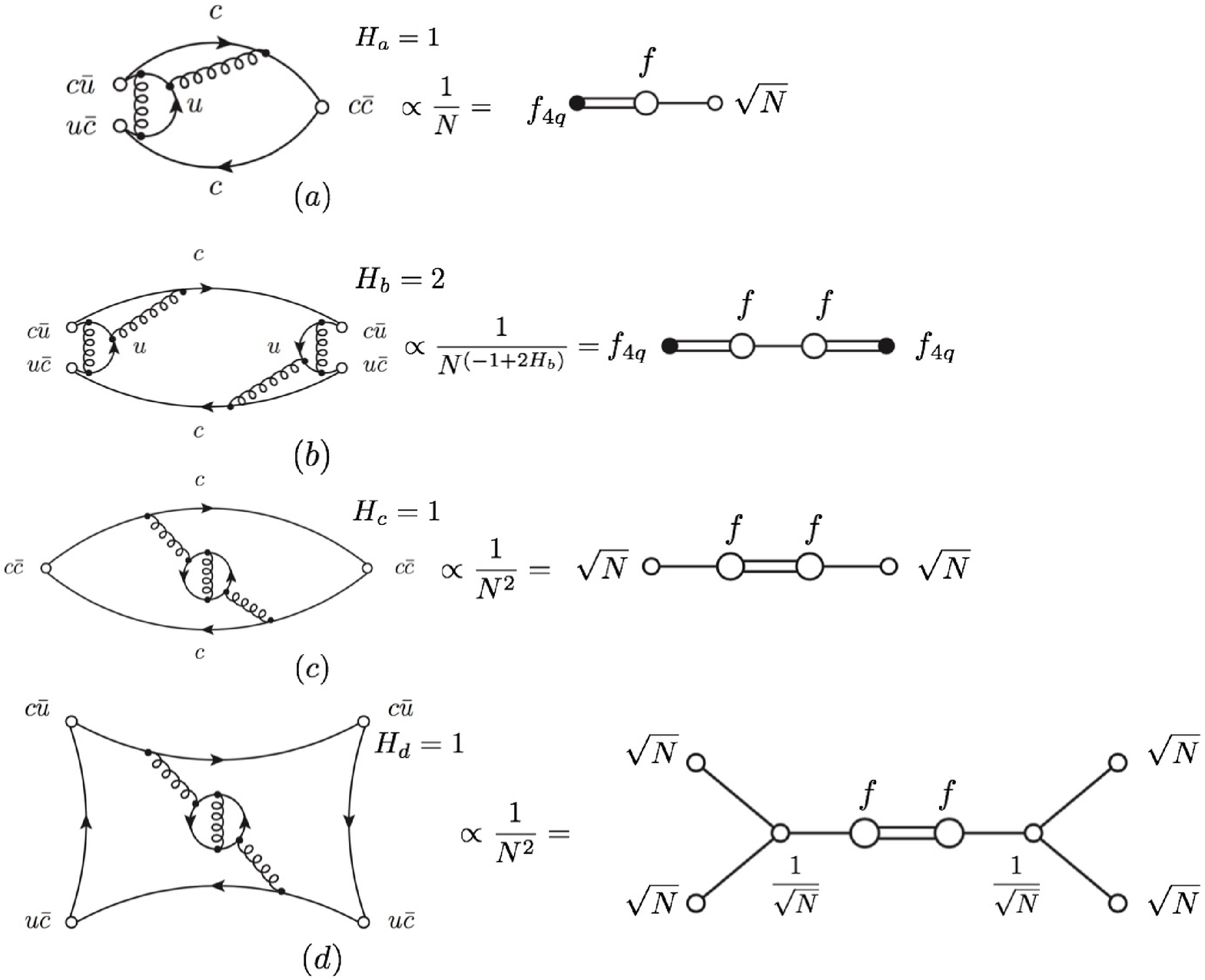}
 \end{center}
\caption{\footnotesize Diagrams describing the mixing of $I=0$ tetraquarks with charmonia. Diagrams (c) and (d) are both non-planar with one handle, see Fig.~\ref{nonplanar}~(c).  Consider for example $(a)$. Here we have $T^a_{ij}T^b_{jr}T^a_{rk}=-1/2N\,T^b_{ik}$ ({\it i.e.} $1/N^2$ with respect to the diagram with no or planar gluons). The diagram is indeed of order $1/N$. To maintain  the same power in the $1/N$ expansion, we cannot shrink $q$ to a point: this is also evident diagrammatically.  That being the case, we will necessarily find a tetraquark in the cut because a gluon, rather than a quark, performs part of the periphery.}   
\label{mixing}
\end{figure}

Here $f_{4q}$ is the amplitude to create the four quark state, the analog of  the previously introduced constant, $f_n$, and $f$ the mixing coefficient (we suppress here possible indices $n$ and $m$ identifying a definite pair of meson and tetraquark). The combination~(\ref{prodf}) appears also in the diagram of Fig.~\ref{mixing} (b) and consistency requires diagram (b) to have two handles. The constant $f$ is found from diagram (c), which describes the process: charmonium$\to$ tetraquark $\to$ charmonium, and which we assume to have one handle only, for simplicity. In total we find the order in $N$ of these constants to be 
\be
f_{4q} \sim N^0,~~ f \sim \frac{1}{N\sqrt{N}}
\label{orderf}
\ee

The diagram of Fig.~\ref{mixing} (d) describes  meson-meson scattering occurring due to the mixing. The three meson vertex is of order $1/\sqrt{N}$, see e.g.~\cite{witten}, and we find
\be
A_{\rm mix}\sim \frac{1}{N^4}
\label{amix}
\ee
In all diagrams of Fig.~\ref{mixing} infinite sums over tetraquark and meson intermediate states are implied. These give rise to contact terms which for brevity are not reported in the diagrams.

The decays into mesons of tetraquarks with $I=0$ occur by two different amplitudes. The amplitude corresponding to the irreducible diagrams analogous to those reported in Fig.~\ref{decay}, with the same $N$ dependence 
\be
g^{(0)}\sim g^{(1)}=\frac{1}{N\sqrt{N}}
\ee
and via the mixing, with  amplitude $g_{\rm mix}$
\be
g_{\rm mix}=\sqrt{A_{\rm mix}}= f~\frac{1}{\sqrt{N}}=\frac{1}{N^2}
\ee

Thus, modes where tetraquarks decay  like charmonium are subleading, which may be not far from reality: after all, the exoticity of $X(3872)$ has been inferred from the fact that it does not decay like a charmonium.

An interesting case is the decay $Y(4260)\to \mu^+ \mu^-$, which is implied by the direct production of $Y(4260)$ in  $e^+ e^-$ annihilation~\cite{Ablikim:2013dyn}. This decay cannot occur to lowest order in $\alpha$ via the irreducible diagrams, as observed in~\cite{dipole}, since the e.m. current can annihilate only one quark-antiquark pair. A sizeable $I=0$ component  in the $Y(4260)$ wave function can give rise to this decay via mixing to the expected (but not yet identified) $L=2, S=J=1$ charmonium.

\section{Meson interactions in the $1/N$~expansion}

For $N$ strictly infinite, $q\bar q$ mesons generated by quark bilinear correlators are free particles. Interactions are generated by letting $N$ to be large but finite. Irreducible vertices with $k$ external mesons are of order $N^{1-\frac{k}{2}}$. Three-meson vertices are of order $1/\sqrt{N}$, quartic meson vertex of order $1/N$. The amplitude of the simplest process, meson-meson  scattering, is of order $1/N$, Fig.~\ref{level} ($b_1$). 

Quark diagrams in the planar approximation generate $q\bar q$ intermediate states, producing the pole terms in the meson diagrams. In the leading $1/N$ approximation, Fig.~\ref{level} ($a_1$), there is only one quark diagram, which contains both $s$- channel and $t$-channel meson poles. In correspondence, the meson-meson amplitude satisfies the DHS duality relation~\cite{DHS} i.e. the sum over $s$-channel resonances reproduces also the $t$-channel resonances, as indicated in Fig.~\ref{level} ($b_1$) where only the sum over the former resonances is reported.


\begin{figure}
 \begin{center}
   \includegraphics[width=0.9 \columnwidth]{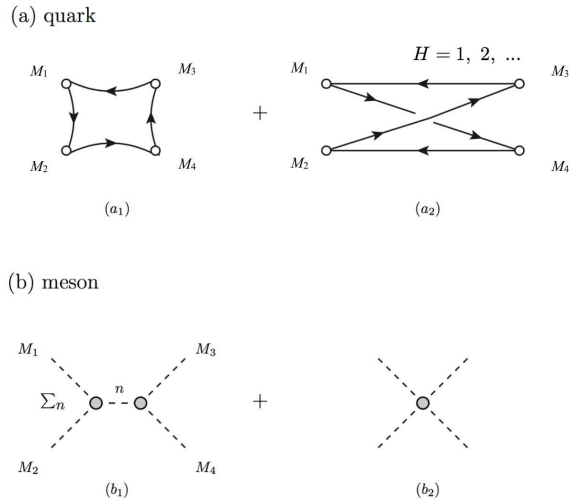}
 \end{center}
\caption{\footnotesize  ($a_1$) leading order quark loop diagram; multiplying at each vertex by the meson normalization factor $1/\sqrt{N}$ leads to an irreducible amplitude of order $1/N$; ($a_2$) non leading order diagrams with handles may contribute additional, tetraquark, poles, as dicussed in the text; ($b_1$) leading order meson-meson amplitude, both sides of the equation $(a_1)=(b_1)$ are of order $1/N$; ($b_2$) additional meson-meson amplitude corresponding the the quark diagram ($a_2$).}   
\label{level}
\end{figure}

We consider first $S$-wave scattering in the channel with $J^P=1^+$ and $G=-1$, {\it i.e.} the quantum numbers of the operator $X^+$ in Eq.~(\ref{xx}).  We have two meson pairs as initial or final states, namely 

\bea
&& y^a= i\frac{( \psi \wedge \rho^+)^a}{\sqrt{2}}      
\label{hidchx}\\
&& z^b= \frac{({\bar D}^0 D^{\star +}-D^+ {\bar D}^{\star 0})^b}{\sqrt{2}} \label{opchx}
\eea
($a$ and $b$ are spin indices) and consider the four possible reactions: $y,z \to y,z$. 

With the chosen flavor composition, the quark loop diagram in Fig.~\ref{level} ($a_1$) contributes only to $z \to z$, where it entails annihilation and recreation of a $c\bar c$ pair and may be neglected, in the limit of large $c$ quark mass. We remain with the second quark loop diagram, Fig.~\ref{level} ($a_2$), giving rise to the meson interaction term in diagram $(b_{2})$. 

Non planar quark diagrams, as in Fig.~\ref{nonplanar}, would develop tetraquark poles in the $s$-channel contributing an amplitude of order $[g^{(1)}]^2=1/N^3$.   

In the simplest approximation in which we neglect final state strong interactions of the color single states, the amplitude for $M_1+M_2 \to M_3 + M_4$ is computed by inserting the spin matrix of each meson (written as $\bar q \,\Gamma_i\, q^\prime)$ in the vertices of Fig.~\ref{level} ($a_2$).We obtain the spin factor~\footnote{ S(12;34)  is the analog of the Chan-Paton factors of dual models~\cite{Paton:1969je}; gamma matrices associated to the quark gluon interaction in Fig.~\ref{nonplanar} (a) do not count since they reduce to $\gamma^0=\pm 1$, in the non-relativistic limit.}
\be
S(12;34)={\rm Tr}\left(\Gamma_1\Gamma_3^\dagger \Gamma_2 \Gamma_4^\dagger\right)
\label{cpfac}
\ee

The quark diagram is such that it transforms the hidden charm (\ref{hidchx}), into the open charm channel, (\ref{opchx}) and viceversa. If we take $M_1 M_2=y$ and $M_3 M_4= z$, we obtain
\bea
&&T(y^a \to z^b) =\frac{i\epsilon^{ade}}{8} \left\{ {\rm Tr}\left[(\sigma^2\sigma^d)(\sigma^2)^\dagger(\sigma^2\sigma^e)(\sigma^2 \sigma^b)^\dagger \right]\right.  -\nonumber \\
&&-\left. {\rm Tr}\left[(\sigma^2\sigma^d)(\sigma^2\sigma^b)^\dagger (\sigma^2\sigma^e)(\sigma^2 )^\dagger \right] \right\}=-\delta^{ab}\nonumber=\\
&&=T(z^b \to y^a)
\label{sfact}
\eea
and
\be
T(y^a \to y^b)=T(z^a \to z^b)=0 \nonumber
\ee

With the S-matrix: $S=1+iT$, we see that, in correspondence to the diagram Fig.~\ref{level} $(a_2)$, the eigenstate of the $J^P=1^+$, $G=-1$ coincide with the combinations $y\pm z$ that are the Fierz rearranged combinations of the symmetric and antisymmetric tetraquark operators, Eq.~(\ref{xx}).
 
It is not difficult to see that the eigenstates of the  $J^P=1^+$, $G=+1$ channels are similarly given by the Fierz rearranged combinations in Eqs.~(\ref{zz},\ref{zzp}).  

Should the quark loop diagram in Fig.~\ref{level} $(a_2)$  develop a pole in one of the eigenchannels, the meson pairs coupled to the resonance would have {\it precisely  the right quantum numbers} to arise from the color Fierz rearranged (symmetric or antisymmetric) diquark-antidiquark states (we stress again that this holds in our very simple approximation of neglecting strong interaction renormalizations of color singlet states arising from the Fierz rearrangement). 

In view of the attractive force indicated in Table~\ref{casimir}, it is tempting to assume that the three channels corresponding to the antisymmetric diquark develop a pole, in which case resonances/molecules at meson-meson thresholds and  tetraquarks would dynamically coincide.

There are two positive aspects of this proposal. 

The first is that the number of resonances thus found is {\it one-half} of what it would be if there were  one resonance for each meson-meson threshold. Stated more explicitly, we seem to see only one and the same resonance in the $J/\psi +\pi^+$ and in the $D +\bar D^\star$ channels, consistently with (\ref{zzmesons}), and not two different ones.

Secondly, for $N\to \infty$ the meson-meson scattering amplitudes vanish and we go back to a free meson theory. In other words meson molecules at threshold have to disappear in the large N limit. 

The situation is similar to what happens for electrons and protons in QED: they are free particles for $\alpha= 0$ but form bound states for $\alpha \neq 0$. Consistency of these two facts requires that the mass of the (hydrogen) bound state is close to the sum of the masses of the free particles:
\be
m_H=m_e + m_p + O(\alpha)
\ee
One could expect a similar mass relation for tetraquarks, {\it e.g.} for the $X(3872)$
\bea
&&m_{X(3872)}=m_{J/\psi}+m_\rho +  O(1/N^p) =\nonumber \\
&&= m_{D} + m_{D^\star} +  O(1/N^p)
\eea
($p\geq 3$) reminiscent of the molecular speculations.

We have to observe that in equation $(\ref{xxmesons})$ the masses of the $\psi \,\rho$ and $D\, D^\star$ thresholds coincide almost exactly  whereas this degeneracy, which may be accidental, is lifted in~(\ref{zzmesons}) and~(\ref{zzpmesons}).

\section{Conclusions}

The correlation functions of antisymmetric and symmetric tetraquark operators are analysed in the large $N$ expansion, with $N$ the number of colors. Unlike what assumed in previous works, we argue that tetraquark poles may emerge only  in non-planar diagrams of the $1/N$ expansion. Combinatoric rules for amplitudes with handles provide a different expectation for the total widths of tetraquarks states, which are found to be $\Gamma\sim 1/N^3$ instead of $\sim 1/N$. Some inconsistencies on the order of the $1/N$ expansion of tetraquark decay constants, $f_{4q}$,  and of  tetraquark-meson mixing amplitudes $f$ are also resolved. 

Starting from this, we have examined the charged eigenchannels of hidden-charm to open-charm meson-meson scattering amplitudes finding a perfect matching with the corresponding Fierz rearranged 
 diquark-antidiquark intermediate states. 

The main outcomes of this analysis are: $i)$ tetraquark correlations have the right degrees of freedom to describe exotic resonances in meson-meson amplitudes $ii)$ in the language of hadron molecules, the number of expected states gets halved when making the connection to diquark-antidiquark antisymmetric states (there is no $X^\prime\sim J/\psi\,\rho^0$ loosely bound molecule). 
The quasi-degeneracy with meson thresholds might have a function at enhancing  tetraquark poles~\cite{fere}.

From this standpoint, selection rules on the spectrum of states may be deduced.  Baryon-meson amplitudes, in the $1/N$ expansion, might eventually shed new light on pentaquark states.

\section*{Acknowledgements} We thank M. Bochicchio, G.~C.~Rossi  and  G.~Veneziano for interesting discussions. Exchanges with A. Ali and a correspondence with C. Hanhart, F-K. Guo, Q. Wang U-G. Meissner and Z. Qiang and with M.~Knecht and S.~Peris are also acknowledged.

\newpage

\end{document}